%% file: main.tex
\documentclass{IEEEtran}
\usepackage{cite}
\usepackage{amsmath,amssymb,amsfonts}
\usepackage{graphicx}
\usepackage{textcomp,nicefrac}
\usepackage{subfig}
\usepackage{float}
\usepackage{color}
\usepackage{booktabs}
\usepackage{multirow}
\usepackage{graphicx}
\usepackage{enumitem}

\usepackage{hyperref}

\def\BibTeX{{\rm B\kern-.05em{\sc i\kern-.025em b}\kern-.08em
T\kern-.1667em\lower.7ex\hbox{E}\kern-.125emX}}
\IEEEoverridecommandlockouts
\IEEEpubid{\makebox[\columnwidth]{978-1-5386-5541-2/18/\$31.00~\copyright2018 IEEE \hfill} \hspace{\columnsep}\makebox[\columnwidth]{ }}

\begin{document}

\title{{Characterizing a Neutron-Induced Fault Model for Deep Neural Networks}}
\author{Fernando Fernandes dos Santos, Angeliki Kritikakou, Josie E. Rodriguez Condia, Juan-David Guerrero-Balaguera\\Matteo Sonza Reorda, Olivier Sentieys, and Paolo Rech
\thanks{
Fernando Fernandes dos Santos, Angeliki Kritikakou, and Olivier Sentieys are with University of Rennes, INRIA, France. E-mail: fernando.fernandes-dos-santos@inria.fr, angeliki.kritikakou@irisa.fr, and olivier.sentieys@inria.fr.
}
\thanks{
 Josie E. Rodriguez Condia, Juan-David Guerrero-Balaguera, and Matteo Sonza Reorda are with Politecnico di Torino - Department  of Control and Computer Engineering DAUIN. E-mail: \{josie.rodriguez, juan.guerrero, matteo.sonzareorda\}@polito.it
}
\thanks{P. Rech is with University of Trento - Department of Industrial Engineering. E-mail: paolo.rech@unitn.it}
}

\maketitle
 \IEEEpubidadjcol
 
\begin{abstract}
\input{src/abstract.tex}

\end{abstract}



\input{src/introduction.tex}
\input{src/background.tex}

\input{src/methodology.tex}
\input{src/beam.tex}

\input{src/faultinjection.tex}
\input{src/conclusion.tex}

\section*{{Appendix A: Research Artifacts}}

{In order to allow the reproducibility of the experiments performed in this paper, we have uploaded the source codes of our setups to online repositories on GitHub. The source code of our beam experiment setup is available at \url{https://github.com/radhelper}. Additionally, we have forked the NVBitFI repository at \url{https://github.com/fernandoFernandeSantos/nvbitfi} to include the FP16, Tensor Cores, and warp-wide fault models. A pull request was submitted on the official NVIDIA NVBitFI GitHub (\url{https://github.com/NVlabs/nvbitfi/pull/19}) to include the new fault models.}

\section*{Acknowledgment}
This project has received funding from the European Union's Horizon 2020 research and innovation
program under the Marie Sklodowska-Curie (MSCA) grant agreement No 886202, from The Coordena\c{c}\~{a}o de Aperfei\c{c}oamento de Pessoal de N\'{i}vel Superior - Brazil (CAPES) - Finance Code 001, from BIENVENÜE Bienvenue Fellowship MSCA Co-funded program, and ANR RE-TRUSTING (ANR-21-CE24-0015-02). 
Neutron beam time was provided by ChipIR (DOI:10.5286/ISIS.E.RB2200004-1, 10.5286/ISIS.E.RB2000137-1, 10.5286/ISIS.E.101136531) thanks to Chris Frost, Carlo Cazzaniga, and Maria Kastriotou and by LANSCE thanks to Steve Wender and Gus Sinnis.

\bibliographystyle{IEEEtran}
\bibliography{IEEEabrv,refreview1}

\end{document}

%% file: src/abstract.tex
The reliability evaluation of Deep Neural Networks (DNNs) executed on Graphic Processing Units (GPUs) is a challenging problem since the hardware architecture is highly complex and the software frameworks are composed of many layers of abstraction. While software-level fault injection is a common and fast way to evaluate the reliability of complex applications, it may produce unrealistic results since it has limited access to the hardware resources and the adopted fault models may be too naive (i.e., single and double bit flip). Contrarily, physical fault injection with neutron beam provides realistic error rates but lacks fault propagation visibility. This paper proposes a characterization of the DNN fault model combining both neutron beam experiments and fault injection at software level. We exposed GPUs running General Matrix Multiplication (GEMM) and DNNs to beam neutrons to measure their error rate. On DNNs, we observe that the percentage of critical errors can be up to 61\%, and show that ECC is ineffective in reducing critical errors. We then performed a complementary software-level fault injection, using fault models derived from RTL simulations. 
Our results show that by injecting complex fault models, the YOLOv3 misdetection rate is validated to be very close to the rate measured with beam experiments, which is 8.66$\times$ higher than the one measured with fault injection using only single-bit flips.

%% file: src/introduction.tex
\section{Introduction}
\label{sec:introduction}

Deep neural Networks (DNNs) are efficient tools to perform several tasks, such as object detection, image segmentation, classification, and prediction~\cite{resnet200, fasterrcnn2015,gkioxari2019}. 
DNNs have been deployed in safety-critical and mission-critical domains, such as robotics, aeronautics and space exploration, smart healthcare, and autonomous driving.
To reduce the processing time and energy consumption of DNN implementations, dedicated architectures with specialized hardware or reduced precision, e.g., half-precision floating-point (16 bits) or even short integer (8 bits), have been proposed, with satisfactory classification, segmentation, and detection accuracy. 
{As an example, Modern Graphics Processing Units (GPUs) feature Tensor Cores, i.e., a dedicated hardware to perform 4$\times$4 matrix multiplications with a single instruction}~\cite{nvidia_volta_2019}. Also, for specific domains, ASIC accelerators are employed for DNN, achieving significant performance and low power consumption~\cite{seventyAuthors}. 

When DNNs are deployed in safety-critical applications, real-time execution and reliability are essential. Thus, to leverage the benefits of DNNs towards safety-critical systems, a DNN-based system must be compliant with standards, such as ISO 26262 and ISO/PAS 21448. %
However, system reliability can be undermined by several sources, including environmental perturbations, aging, and process/temperature/voltage variations~\cite{laprie95,nicolaidis99}. Radiation-induced soft errors are particularly critical, as they dominate error rates in commercial devices~\cite{Baumann2005} and are very challenging due to their random and transient nature~\cite{Muh05}. As a consequence, DNNs are susceptible to transient faults induced by radiation~\cite{Santos_TR,SCHari2017DNNs,  Ibraim2020, edgeAI2021, coralRubens2022}. Additionally, GPUs have a high fault rate because of the high amount of available resources~\cite{Nathan2014,Tiwari2015, Sulivan2021NVIDIADDR} and the possibility to have multiple output elements corrupted, which undermines DNN reliability~\cite{Santos_TR, Ibraim2020}.

The reliability evaluation of GPU architectures is then mandatory to characterize the errors that can hinder the DNN accuracy. A realistic method to evaluate the reliability of DNNs on GPUs is physical-level fault injection with a neutron beam. However, with beam experiments, only the fault outcome in the application's output can be observed. Thus, fault propagation visibility is missing. Contrarily, fault injection at the software level has limited overhead, making it possible to perform several fault injection campaigns in order to identify the faults that impact the DNN reliability. Despite being fast, software-level fault injection is restricted by the resources where faults can be injected, such as register file and variables, limiting the accuracy of the fault models. 
The main goal of this work is to bridge this gap by obtaining a realistic DNNs' transient error model, extracted from neutron beam experiments, and compare it with different error models, obtained from software-level fault injections.

\begin{figure*}[ht]%
    \centering
    \subfloat[Expected object detection]{
        \includegraphics[width=0.32\textwidth,keepaspectratio]{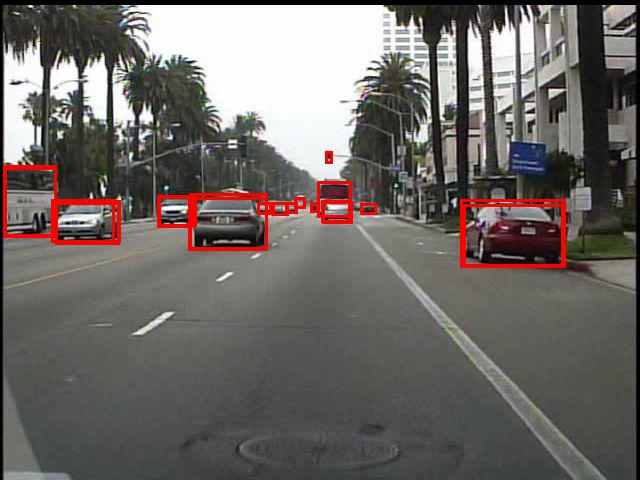}
        \label{fig:gold_det}
    }%
    \subfloat[Critical SDC from radiation experiments]{
        \includegraphics[width=0.32\textwidth,keepaspectratio]{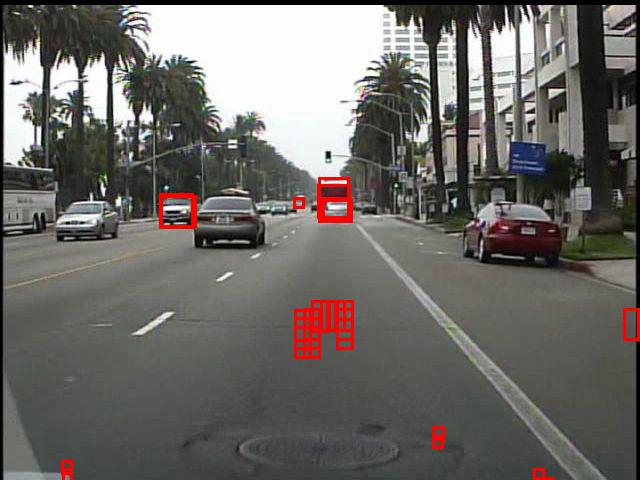}
        \label{fig:rad_det}
    }%
    \subfloat[Critical SDC from NVBitFI fault injection]{
    \includegraphics[width=0.32\textwidth,keepaspectratio]{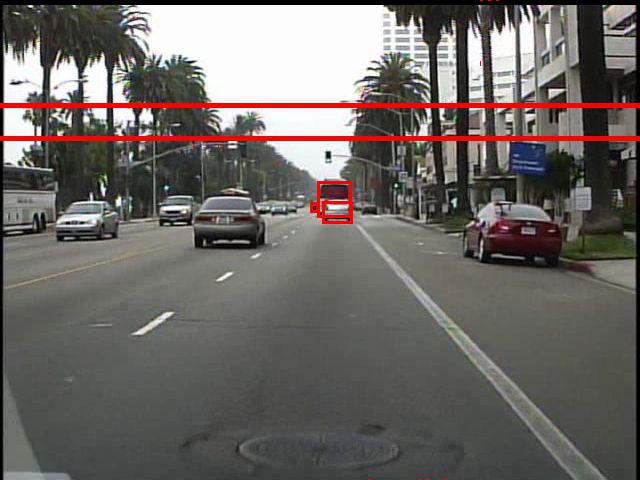}
    \label{fig:fi_det}
    }%

    \caption{Figure~\ref{fig:gold_det} shows the expected outcome from the YOLOv3 DNN when no fault is observed. Figures~\ref{fig:rad_det} and~\ref{fig:fi_det} show the object detection when a fault perturbates the execution of the DNN from neutron beam experiments and software-level fault injection, respectively. Both Figures~\ref{fig:rad_det} and~\ref{fig:fi_det} are Critical SDCs. The fault model that generated Figure~\ref{fig:fi_det} is based on the proposed complex fault model. The frame is extracted from Caltech Pedestrian detection dataset~\cite{caltechDollar2009}.}%
    \label{fig:case}
\end{figure*}

We report findings from beam testing campaigns that assess GPU reliability as a DNN accelerator and identify the causes of \textit{critical} errors (i.e., errors that impact detection) and \textit{tolerable} errors {(i.e., errors that do not impact the detection)}. 
Our testing campaigns consider two different GPU architectures, i.e., Kepler (Tesla K40) and Volta (Tesla V100), and several configurations for multiple DNNs (YOLO, Faster R-CNN, ResNet). These configurations include different floating-point precisions and the use or non-use of specialized cores and of Error Correction Codes (ECC). We also evaluated the error rate of the fundamental operations of DNNs, the General Matrix Multiplication (GEMM) kernel. 
The results show that a single particle can spread through the GPU microarchitecture, affecting several parallel threads and output elements, significantly impacting DNN reliability.
Furthermore, for DNNs, Single Error Correction Double Error Detection (SECDED) ECC can be useful, but not always effective. 
{
Although the ECC can reduce the error rate by an order of magnitude, ECC can reduce neither the number of critical nor multiple errors that come from unprotected GPU units. We show that even with ECC protection, the critical error rate is not neglectable.
}

We modify the NVIDIA Binary Instrumentation Fault Injector (NVBitFI)~\cite{nvbitfi2021} to evaluate the fault propagation for complex fault models on GEMM and DNNs. With NVBitFI, we are able to mimic the critical errors observed in beam experiments. We inject 2,000 faults on YOLOv3 DNN and 644 faults on GEMM per fault model for each configuration, accounting for more than 35,000 injected faults. The contributions of this paper are as follows:
\begin{itemize}
    \item {
    As we perform the experiments with an accelerated neutron beam (with a flux $\approx8\times$ higher than the terrestrial flux), we provide a realistic error rate that accounts for more than 380,000 years of terrestrial device operation.} We show the results for two different GPU architectures (Kepler and Volta), with different floating-point precisions (FP32 and FP16).
   \item We measure the efficacy of the ECC in protecting the DNNs against errors that impact the classification/detection. We also characterize the criticality of the errors that modify the GEMM output when used as the core of the DNNs operations.
    %
    %
    \item We present a complete characterization of the fault model at the software level on GEMM and DNNs. On this fault propagation evaluation, we can identify which types of simulated faults affect the DNN output similarly to the beam experiments.
    \item We craft an improved version of NVBitFi that includes multiple thread fault models and compare them with the standard single-bit flip.
\end{itemize}

The rest of this paper is organised as follows. Section \ref{sec:background} presents the background on radiation-induced errors on DNNs. Section~\ref{sec:methodology} introduces the proposed analysis and shows the experimental methodology for the beam experiments and the fault simulation at the software level. The GEMM and DNNs error rate is presented in Section~\ref{sec:fit_beam}. Section~\ref{sec:fault_injection} presents the fault injection results at the software level using the proposed fault models on YOLOv3 with FP16 and FP32. Finally, Section~\ref{sec:conclusions} concludes the paper.

%% file: src/background.tex
\section{Radiation induced errors in DNNs}
\label{sec:background}


{
Terrestrial neutrons can interact with electronic devices, and a transistor’s state can be perturbed by a neutron strike (Single Event Transient, SET), thus generating bit-flips in memory or current spikes in logic circuits that, if latched, may lead to an error~\cite{Mahatme11} (i.e., Single Event Upset, SEU). An error generated by a neutron is called a soft error, as most of the time, it does not damage the device,} but it can change the output of an application (e.g., a DNN classification). On GPUs, as with any other device, the fault propagates from the hardware to the software level leading to the possible outcomes:
\begin{enumerate}
\item \textbf{Masked}: no effect on the program output. The corrupted data is not used, or the circuit functionality is not affected. 
\item \textbf{Detected Unrecoverable Error (DUE)}: the program stops working or the entire system crashes.
\item \textbf{Silent Data Corruption (SDC)}: undetected output corruption, that is, the application finishes, but the application output is not correct and no flag or indication of an error is set. 
\end{enumerate}

\begin{figure*}[t]
    	\centering{
        \includegraphics[width=0.9\linewidth]{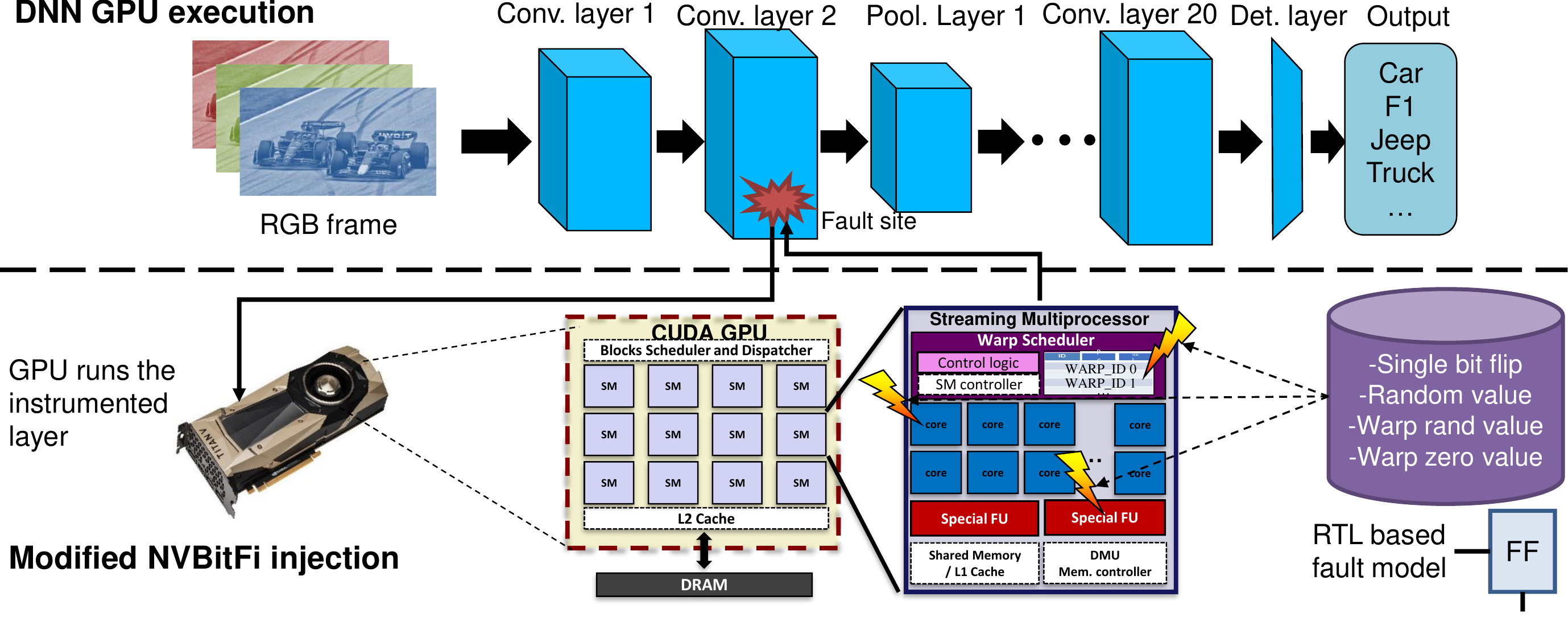}
        \caption{Proposed software-level fault approach. The DNN receives an RGB frame in the first convolution and propagates it into the subsequent layers. In the modified version of the NVBitFi, the kernels are instrumented, and a random fault site inside the layer kernel is selected. Using the fault models based on~\cite{SantosEstebanDSN2021}, we can inject multiple types of faults, corrupting single registers or one register in an entire GPU warp.}
        \label{fig:idea}
    }
\end{figure*}

The SDCs on DNNs have a particular characteristic compared to other typical GPU applications. The SDCs can be separated into two classes: (1) \textbf{Tolerable SDCs} are the errors that change the output of the last layer of the DNN, but they do not modify the classification or object detection. (2) \textbf{Critical SDCs} are the errors that change the classification or detection partially or entirely. 
The Critical SDCs are particularly dangerous for safety-critical applications, as a decision based on an incorrect inference can lead to a catastrophic event. For instance, Figure~\ref{fig:case} shows the result based on a real scenario extracted from Caltech Pedestrian detection dataset~\cite{caltechDollar2009}. The expected object detection is shown in Figure~\ref{fig:gold_det}. Figures~\ref{fig:rad_det} and~\ref{fig:fi_det} show two examples of Critical SDCs observed on neutron beam experiments and software-level fault injection, respectively. 
{
In fact, critical SDCs can be detected at the software level with different types of fault tolerance, such as temporal redundancy~\cite{Draghetti2019, Saurabh2022}, algorithm-based fault tolerance~\cite{Santos_TR, Long2019, Mittal2020, Sulivan2021NVIDIADDR, Ibraim2020}, and float point value clipping~\cite{Santos_TR, SC16}. However, all the software-level approaches add overhead to the DNN inference. Contrarily, we propose a better characterization of the fault model of the DNNs. Then researchers can use more efficient ways to detect faults, such as fault-aware training to harden the DNN from the design phase, which is only possible by knowing the correct fault model of the accelerator beforehand.
}

Different types of fault injection are available to evaluate the reliability of GPUs, such as physical-level fault injection with beam experiments and fault simulation at the software level. In previous works, the GPU error rate for typical applications has been evaluated through beam experiments where the device is exposed to a particle flux and the error rate is measured~\cite{tc2016, Santos_TR, luisEntrena2020Sensitivity, dueKojiro2021, luisEntrena2021Protons, Sulivan2021NVIDIADDR}.
When the accelerated particle beam goes through the hardware, it induces transient faults, which may lead to output errors. The main characteristic of neutron beam experiments is that the whole device is irradiated, and thus, faults are not injected only to a subset of resources. 
To measure the neutron-induced error rate of DNNs, researchers exposed GPUs~\cite{Santos_TR, Ibraim2020}, FPGAs~\cite{Libano2021TNS}, and ASIC accelerators~\cite{coralRubens2022}, running various types of DNNs.
Beam experiments provide a realistic error rate of the device running a code. However, we cannot distinguish which level of the hardware or the software contributes more to the device error rate, making it challenging to identify the most vulnerable parts of the system. Knowledge of fault propagation through the different hardware and software layers would help system designers to improve reliability more effectively.

In order to understand the faults' propagation, fault injection is performed by simulation. Faults can be injected at different levels of abstraction, from low-level Register-Transfer Level (RTL)~\cite{Iturbe2016,Cho2013,condia2020} to microarchitecture~\cite{Chatzidimitriou2017, Constantinescu2012} and software~\cite{Ferraretto2016, GPUQin, PytorchFIMahmoudAggarwalDSML20, softwareResilienceEvgenia2021, nvbitfi2021, Bolchini2022DNNFaultModel}. Each level of abstraction provides the propagation probability of the injected fault to the output, measured in terms of Hardware, Architectural, Software, or Program Vulnerability Factor (HVF, AVF, SVF, PVF, respectively)~\cite{Mukherjee2003, vulnerabilityDimitrisISCA2021}.
Fault injection assumes that the fault already occurred and provides the fault propagation through the system. However, this approach has limitations, since the fault model and fault injection probabilities are defined by the user and the simulator, which can lead to unrealistic results. Moreover, faults can be injected only into the available and accessible resources, which is a subset of the resources and not the entire system, as with beam experiments.

In recent works, the DNNs fault model on GPUs has been analyzed to provide a more realistic reliability evaluation using fault injection~\cite{Santos_TR, SantosEstebanDSN2021, Bolchini2022DNNFaultModel}.
The fault model analysis is based on the key concept of matrix multiplications for DNNs. Most DNN layers, such as convolutional, recurrent, and fully connected, can be represented as General Matrix Multiplication (GEMM). Then, the reliability evaluation can be performed at the GEMM module level. Consequently, the fault impact on the GEMM output is limited to a 2D geometry format. Therefore, once the fault is injected, the error will manifest as a corruption of a single element, line/row, or block of the output matrix.
The main limitation of the recent works regarding realistic fault models is that they are either based on beam experiments, or they perform fault injection in a high level of abstraction (C++/Python), without taking into account the GPU architecture details that impact the fault propagation.

In this work, we leverage the knowledge obtained in the past works with beam experiments~\cite{Santos_TR} and RTL fault injections~\cite{SantosEstebanDSN2021} {
to propose an analysis that considers the fault propagation in GEMM modules with Shader Assembler (SASS) -level fault injections (i.e., assembly level on NVIDIA GPUs).
}

%% file: src/methodology.tex
\section{Evaluation Methodology}
\label{sec:methodology}

This section presents the analysis we propose to characterize the GPU fault model for DNNs and how we modify the NVBitFI to meet our needs. Then, we carefully present the characterized GPU devices, the considered DNNs, the metrics adopted for the reliability evaluation, and how the beam experiments and fault injections are performed.

\input{src/idea}

\subsection{GPU devices}

We consider Kepler (Tesla K40) and Volta (Tesla V100) NVIDIA GPUs. The tested NVIDIA K40~(\textbf{Kepler}) is built with the Kepler ISA and fabricated in a $28nm$ TSMC standard CMOS technology~\cite{nvidia_kepler_2012}. 
Tesla V100 (\textbf{Volta}) are designed with the Volta microarchitecture and built with TSMC FinFET $12nm$~\cite{nvidia_volta_2019}.  Volta GPUs feature hardware acceleration for three {IEEE 754-2008} floating-point point precisions: double (FP64), single (FP32), and half (FP16). We also evaluate the reliability of Volta GPUs' \textit{tensor cores}, i.e., a specific hardware that performs {$4\times4$} Matrix Multiplication from GEMM kernels in one cycle with FP16 precision. Tensor cores can also perform matrix multiplication for FP32 precision, however, the data will be cast to FP16 in the low-level operation.
Both GPU architectures have available Single Error Correction Double Error Detection (SECDED) Error Correcting Code (ECC) to protect the register file, shared memory, and caches. Our evaluation considers errors occurring only in the GPU core, not in the main memory. For {Kepler} devices, we chose a beam spot that is sufficiently small (2cm of diameter) in order to not hit the onboard {DRAM}, when ECC is disabled. {Since Volta GPUs' DRAM is too close to the GPU core, we choose to test Volta GPUs only with ECC enabled.}

\subsection{Deep Neural Networks}

This work considers three modern DNN models: 
i) You Only Look Once (YOLO v1 and v3)~\cite{yolov1, yolov3}, ii) a Faster Region-based Convolution Neural Network (Faster R-CNN)~\cite{fasterrcnn2015}, and iii) a Residual Network (Resnet)~\cite{resnet200}.
\textbf{YOLO} is based on \textit{Darknet}, which is an open-source CNN in C and CUDA~\cite{yolov1, yolov3}. \textbf{Faster R-CNN} is written in C++ and Python, based on the Caffe~\cite{Jia2014} deep learning framework. \textbf{ResNet} is a CNN based on the Torch deep learning framework~\cite{torch7}. ResNet only performs object classification, while YOLO and Faster R-CNN also provide object detection. 

We classify the observed SDCs between Tolerable SDCs (i.e., that do not impact classification/detection) and Critical SDCs (i.e., that impact classification/detection).
When radiation modifies the classification or detection, we check whether all the objects are detected and classified correctly. Then, we check if the error created a false positive (i.e., add non-existing objects) or if the error made the DNN misdetect objects. 
For the beam experiments and the fault injection, we evaluate the reliability of the DNNs with two datasets: Caltech Pedestrian detection~\cite{caltechDollar2009} and VOC2012~\cite{VOCPascalEveringham15}.
Note that, evaluating with fault injection all the frames tested on neutron experiments would be unfeasible as the fault sites grow exponentially. Thus, 
to reduce the fault space for the SASS-level fault injections, we select the frames from the Caltech dataset that generated critical SDCs on the beam experiments. 

\subsection{Beam Experiment Setup}

\begin{figure}[ht]
	\centering{
		\includegraphics[width=0.98\linewidth]{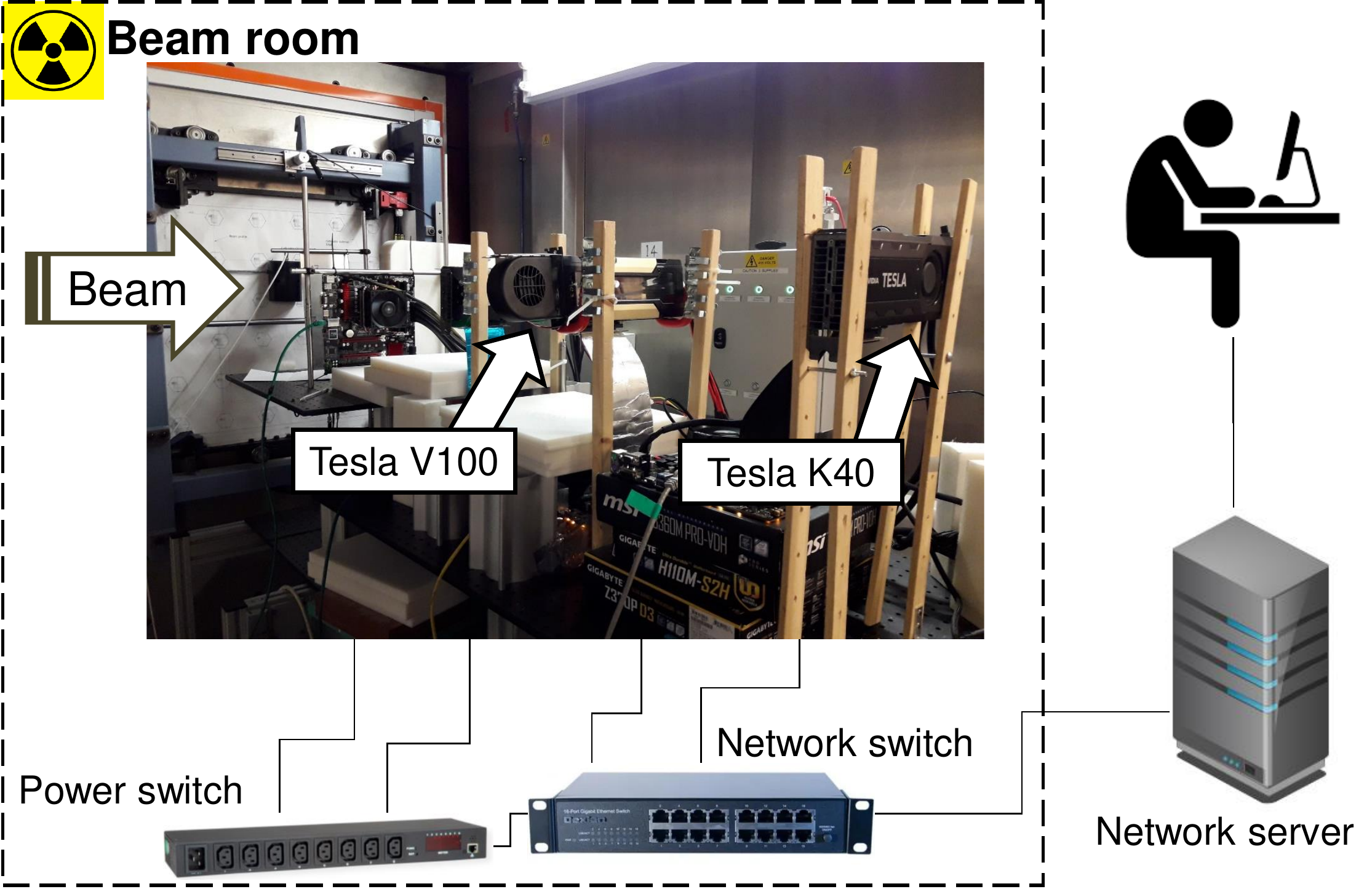}
		\caption[]{Experimental setup at ChipIR at Ruthenford Appleton Laboratory in UK. Boards are placed on the beam room and exposed to a neutron flux of $\approx 3.5\times10^6 n/(cm^{2}/s)$. The network server is composed of a set of Python scripts which controls the devices through the Ethernet and performs power cycles through a power switch.}
		\label{fig:setup}
	}
\end{figure}

Our experiments were performed at the ChipIR facility of the {Rutherford} Appleton Laboratory, UK, and at the LANSCE facility at Los Alamos National Laboratory, US. Figure~\ref{fig:setup} shows the setup mounted in the ChipIR facility. Both facilities deliver a beam of neutrons with a spectrum of energies similar to the atmospheric neutron one~\cite{Cazzaniga_2018}. The available neutron flux was about $3.5\times10^6 n/(cm^{2}/s)$, $\sim$8 orders of magnitude higher than the terrestrial flux (13 $neutrons/(cm^{2} \cdot h$) at sea level~\cite{Jedec2006}). The Failure In Time (FIT) rate is calculated by dividing the number of observed errors by the received particles fluence ($neutrons/cm^{2}$). 

Since the terrestrial neutron flux is low, it is improbable to see more than a single corruption during program execution in a realistic application. We have carefully designed the experiments to maintain this property (observed error rates were lower than 1 error per 1,000 executions). In this way, the experimental data can be scaled to the natural terrestrial environment without introducing artifacts. 
Each DNN was tested for at least 24 effective hours, not including the setup, result check, initialization, and recovery from the DUE time.

We added setup software and hardware watchdogs to monitor the experiments (Figure~\ref{fig:setup}). 
The software watchdog controls the application under test, and if it stops responding in a predefined time interval, the kernel is killed and relaunched. This watchdog detects kernel crashes or software hangs, i.e., application crashes or control flow errors occur that prevent the GPU from completing assigned tasks (e.g., an infinite loop). The hardware watchdog is an Ethernet-controlled switch that performs a host computer's power cycle, if the host computer itself does not acknowledge any ping requests in a predefined time interval. The hardware watchdog is necessary to detect operating system hangs. {The source code of our setup and other research artifacts used in this paper are available in online repositories (details in Appendix A).} 

\subsection{SASS-level fault injection}

To perform the SASS-level fault injection, we have updated the available NVBitFI framework~\cite{nvbitfi2021}. NVBitFI allows instrumenting the GPU kernels at the assembly level (i.e., at the microinstruction level). Other fault injectors, such as GPUQin, CAROL-FI, Kayotee, GPGPU-SIM, SASSIFI~\cite{GPUQin, sc17Daniel, jha2019kayotee, sassifi2017} neither inject at the SASS level nor offer support for Volta and newer architectures, nor inject in CUDA libraries.

NVBitFI can inject transient errors in the GPU's ISA visible states, modifying the SASS instructions output of a code being executed on a real GPU. By default, NVBitFI allows for the users to select the fault model to inject single and double bit-flip, single zero value, and a single random value. We modify the injection procedure to inject the multiple thread faults, Warp random value and Warp zero value. As we used an NVIDIA CUDA Warp primitive to perform the synchronization between the Warp threads, our modifications add a low overhead to the expected NVBitFI injection overhead. Our modified NVBitFI is similar to the original version, but when the Warp-wide injection mode is selected, it will modify one output of one floating-point instruction in all threads within a random warp of the GPU.

%% file: src/idea.tex
\subsection{Proposed analysis}

For the first time, we propose implementing and evaluating different fault models for GPUs at the instruction-level.{
The proposed fault models are based on the fault syndrome observed on RTL fault injection~\cite{SantosEstebanDSN2021}, where the authors reported multiple fault syndromes based on fault injections on a synthetic RTL version of an NVIDIA microarchitecture GPU~\cite{FlexGrip2013, condia2020}.
}
We inject faults on DNNs using an instruction-level fault injector (NVBitFI). With more realistic fault models, we can verify whether the DNN criticality obtained from fault simulation is comparable to the one observed with beam experiments. 
{In fact, the warp-wide fault models were previously proposed by Siva \emph{et al.}~\cite{sassifi2017} in the SASSFI fault injector. However, SASSIFI only supports older NVIDIA architectures, making it impossible to simulate complex applications like DNNs on newer GPU architectures. Additionally, our analysis is not limited to the fault model itself, but also considers the consequences of complex faults on the DNNs inference process and compares them with the results from beam experiments.}

Figure~\ref{fig:idea} shows the proposed SASS level fault injection methodology.
When a GPU kernel is launched, the threads are split into small groups of 32 threads called \textit{Warps}. A Warp is the smallest unit of execution on NVIDIA GPUs, where all threads within a Warp start executing the kernel instructions together and finish simultaneously. As reported in past works~\cite{tc2016, Lunardi2018, condia2020, dueKojiro2021, luisEntrena2020Sensitivity, SantosEstebanDSN2021}, the Warp scheduler and the control logic, responsible for managing the Warp's threads, are critical resources, where a single event upset can generate errors affecting all the 32 Warp threads.
Aiming to model this behavior at the software level, we modify the NVBitFI fault injector to support two new fault models, i.e., Warp random value and Warp zero value. These two fault models are derived from RTL level fault injections~\cite{SantosEstebanDSN2021}, where a single fault on GPU shared resources, such as Warp scheduler and special functional units, is capable of affecting the output of all threads in a GPU Warp.

The modified version of NVBitFI selects a random fault site on the DNN under evaluation. The fault site can be any module of the DNN. Then, we select one of the following fault models to be injected into the output of one or more instruction register(s): (1) standard \textbf{Single Bit Flip} in the output of one instruction (32-bit register); (2) a \textbf{Random value} to replace the value in the instruction output; (3) \textbf{Warp random value}, where all the threads within a warp will have at least one instruction output replaced by a random value; (4) \textbf{Warp zero value}, which is similar to Warp random value, but we replace with a zero value instead of a random one.


%% file: src/beam.tex
\section{DNNs Error Rate}
\label{sec:fit_beam}

\begin{figure*}[ht]%
	\centering
	\subfloat[]{
		\includegraphics[width=0.5\textwidth,keepaspectratio]{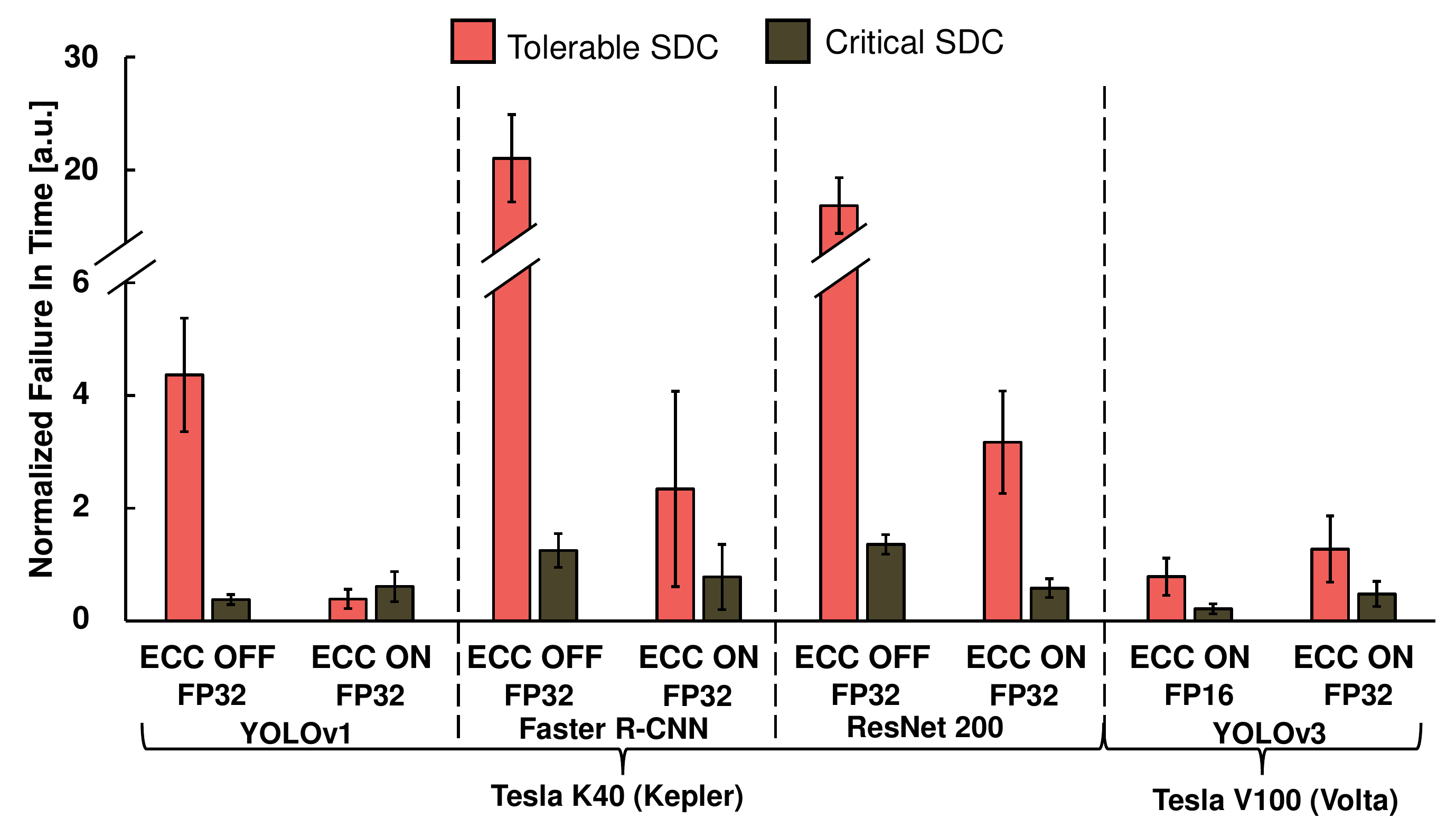}
		\label{fig:fit_sdc}
	}%
	\subfloat[]{
		\includegraphics[width=0.5\textwidth,keepaspectratio]{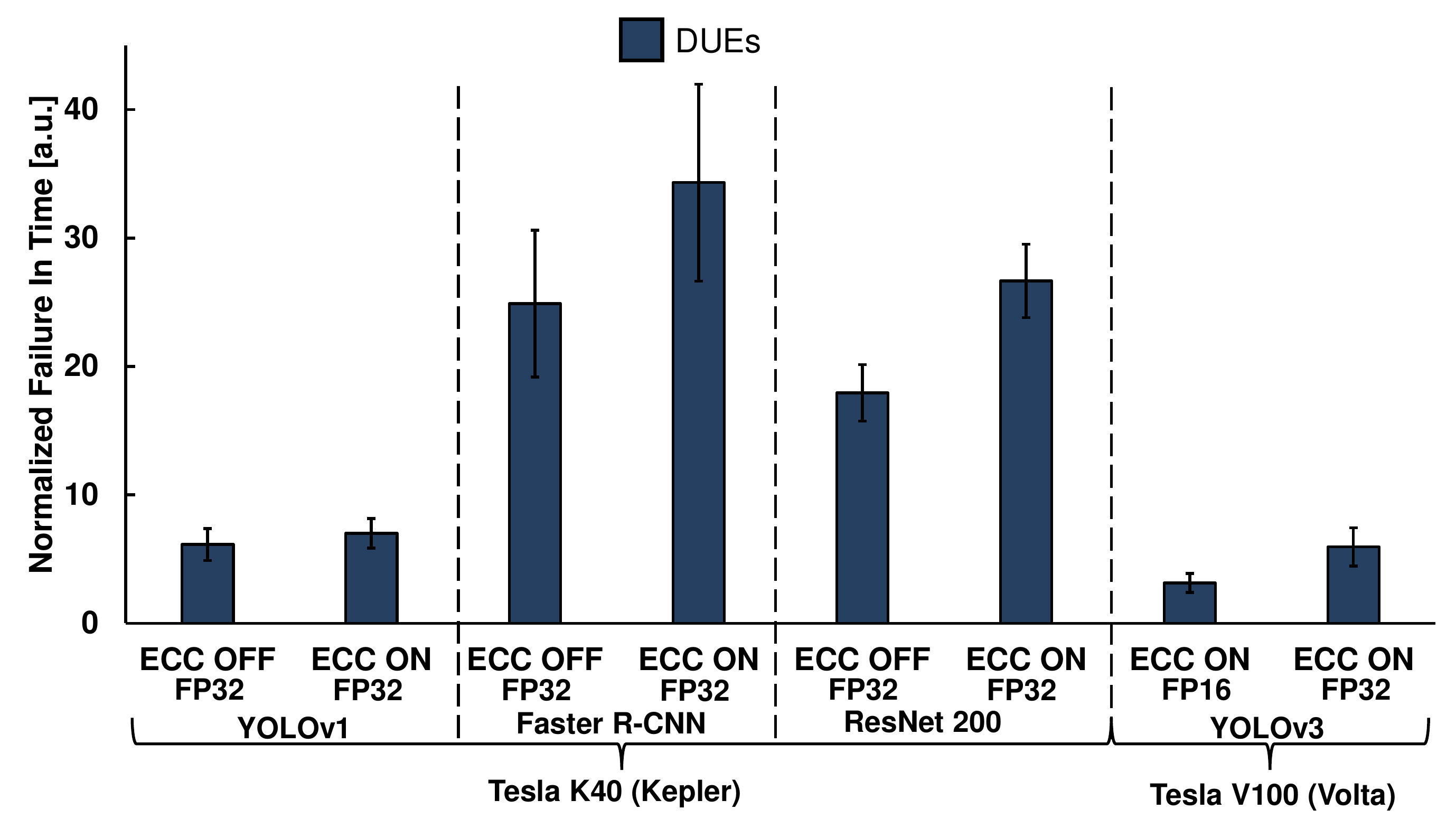}
		\label{fig:fit_due}
	}%
	\caption{{Figure~\ref{fig:fit_sdc} and \ref{fig:fit_due} show the normalized Failure In Time of SDC and DUE, respectively, for YOLOv1, Faster R-CNN, ResNet, and YOLOv3. The SDC rate is separated between Tolerable errors (i.e., the inference is not modified) and Critical SDCs (i.e., the inference is modified).}}%
	\label{fig:fit_all}
\end{figure*}

This section presents the results from radiation experiments for the evaluated Deep Neural Networks (DNNs). To provide insight into the DNNs' error model, we present an evaluation of the error rate and model of GEMM, which is the core of current DNNs. 

\subsection{Deep Neural Networks FIT rate}

{Figures~\ref{fig:fit_sdc} and~\ref{fig:fit_due} show the normalized SDC and DUE rates, respectively. We show results for YOLOv1, Faster R-CNN, and ResNet on Kepler and YOLOv3 on Volta GPUs. We separate the Tolerable and Critical SDCs.} The reported results are normalized to not reveal business-sensitive information. Even if the data is normalized, it allows a direct comparison among configurations. The reported values are relative to  YOLOv1 ECC ON SDC FIT for Kepler and YOLOv3 ECC ON SDC FIT for Volta. Experimental data is presented with a 95\% confidence interval.

{Figure~\ref{fig:fit_all} shows that} DUEs are more probable than SDCs for all tested configurations.
DNN kernels have a high level of reuse that requires several device-host (CPU-GPU) synchronizations. A transient fault during these synchronizations could potentially result in a GPU DUE. As a result, while a significant portion of SDCs could be masked, DUEs could still undermine the device's reliability. For the same reason, Faster R-CNN and Resnet, which require a much larger number of synchronizations, show up to $5\times$ higher DUE rate than YOLO. 

{The impact of the ECC on the GPU’s DUE rate has been evaluated in past works~\cite{Tiwari2015, Lunardi2018}, which show that the DUE rate increases when ECC is ON, particularly in the case of memory-intensive codes. With ECC enabled, the DUE rate increases by up to 30\% for Faster R-CNN, Resnet, and YOLOv1. ECC is able to correct one-bit flip in the protected memories, and, when a double-bit flip is detected, it throws a system exception. Consequently, as DNNs use a large memory to perform classification/detection, multiple errors on memories are expected to happen, resulting in a higher DUE rate when ECC is ON. }
Equivalently, on Volta GPU, the DUE rate grows as the data representation increases. Comparing FP16 and FP32 versions of YOLOv3, the memory size grows $2\times$. 
As the FP32 version of YOLOv3 performs more memory transfers than the FP16 version, the DUE is expected to be higher for FP32. Additionally, with the ECC ON on Volta, and when caches and register usage increases to store the FP32 precision, the double bit flips in the memories are more frequent.

Figure~\ref{fig:fit_sdc} shows that the SDC rates are related to the DNN model complexity and accuracy. Compared to YOLO, the complex structures used on Faster R-CNN and Resnet increase their SDC rate by more than $10\times$. Figure~\ref{fig:fit_sdc} shows that Resnet has a higher FIT rate than YOLO (i.e., v1 and v3), though the rate is similar to Faster R-CNN. Although Faster R-CNN and ResNet have a high detection/classification accuracy, it is insufficient to compensate for the higher error rate associated with the complex framework used for both DNNs. A more accurate DNN, most of the time, demands more computations to be performed. Consequently, as the error rate is directly proportional to the amount of resources used, the more resource demanded, the higher the error rate.

The SDC rate for YOLOv1, Faster R-CNN, and Resnet with ECC ON is $21\%$, $13.6\%$, and $22\%$ less than the SDC rate seen with ECC OFF, respectively. ECC is not as effective in these workloads as in other codes, mainly because neural networks are intrinsically resilient to data errors. ECC can reduce the GEMM SDC rate by about one order of magnitude~\cite{tc2016} (details in Section~\ref{subsec:error_model}). However, ECC is less efficient in protecting DNNs, as some of the SDCs that ECC masks would not have affected the DNN execution. {Figure~\ref{fig:fit_sdc} shows that even if ECC significantly reduces the number of Tolerable SDCs, ECC fails at reducing the occurrence of Critical SDCs.}
On Volta GPU, YOLOv3 FP16 has a lower FIT than YOLOv3 FP32. The amount of per-core resources required to perform the operations, depends on the chosen data precision. Namely, fewer resources will be used for the lower floating-point precision. 
For object detection, the percentage of critical SDCs is much lower for Faster R-CNN and YOLOv3 than for YOLOv1. For YOLOv1, the percentage of critical SDCs is 8\% with ECC OFF and 61\% for the Kepler with ECC ON. For YOLOv3 with ECC ON, the percentage of critical SDCs is 21\% and 27\% for FP16 and FP32, respectively. For Faster R-CNN, the critical SDCs are 5\% with ECC OFF and 25\% for the Kepler with ECC ON. The difference in the percentages of critical errors among DNNs comes from the detection mechanism. YOLOv1 is the simplest DNN tested in our setup with 31 layers. An error is expected to impact much more the detection of simpler DNNs than more complex ones. 

{Although ECC reduces the error rate, the portion of critical errors is not reduced at the same rate as tolerable errors. This can be explained with the fine-grain analysis based on GPU-Qin~\cite{SC16}. As shown in~\cite{SC16}, ECC does not mask all possible faults but only the ones occurring in the protected memories, while faults in computing elements can propagate to the output.}
ECC reduces the absolute number of SDCs, but it has the side effect of increasing the portion of multiple errors (details in Section~\ref{subsec:error_model}), which are more likely to propagate through DNN layers and affect detection. 
On Volta, the percentages of critical errors are similar comparing the two data types when ECC is ON. It is worth noting that critical SDCs are less dependent on data type as the two DNNs have similar detection accuracy regarding the data precision. 

\input{src/complex_errors.tex}

%% file: src/complex_errors.tex
\subsection{Complex error models for DNNs}
\label{subsec:error_model}

\begin{figure*}[ht]%
    \centering
    \subfloat[]{
        \includegraphics[width=0.5\textwidth,keepaspectratio]{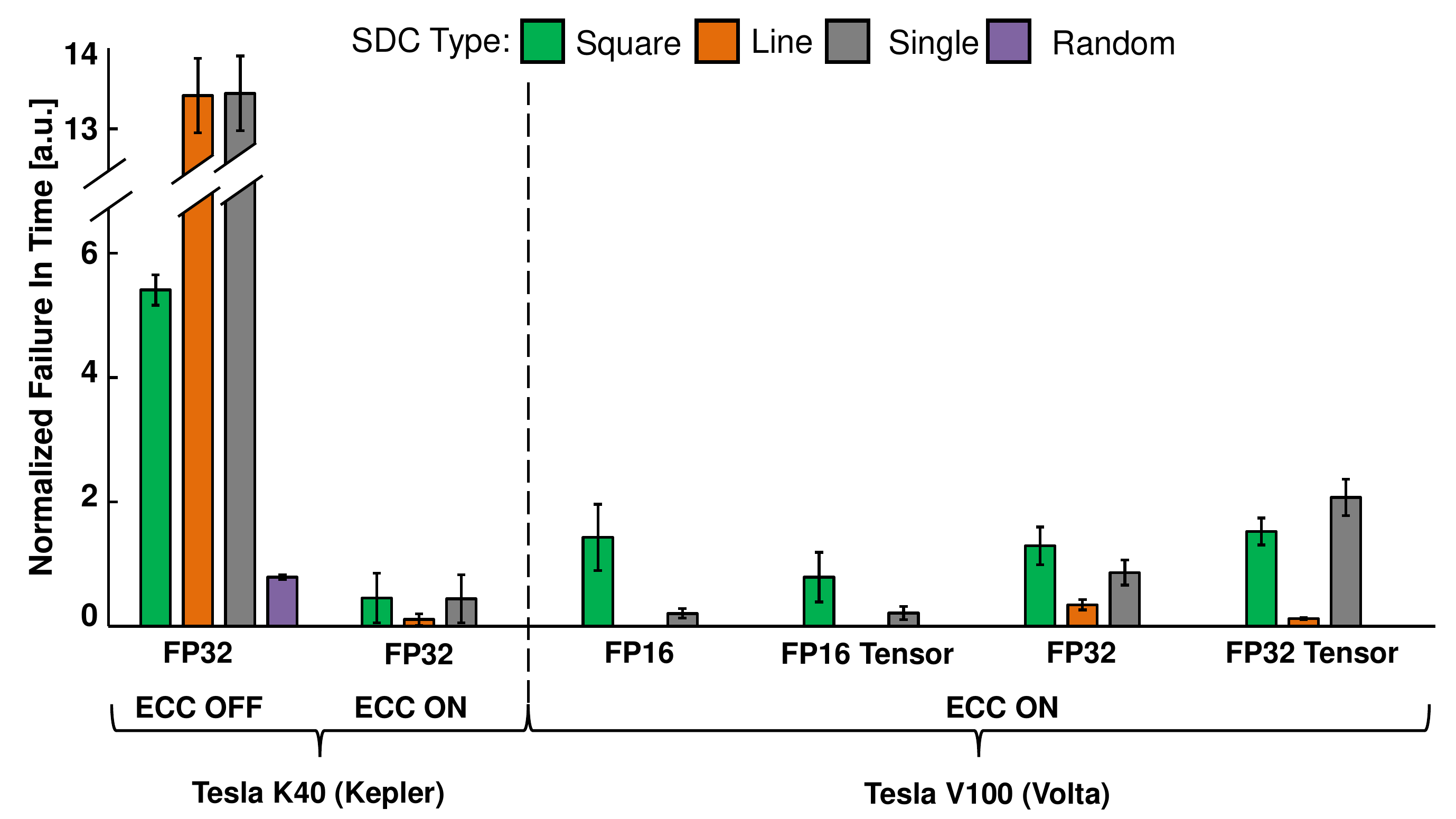}
        \label{fig:fit_gemm_sdc}
    }%
    \subfloat[]{
        \includegraphics[width=0.5\textwidth,keepaspectratio]{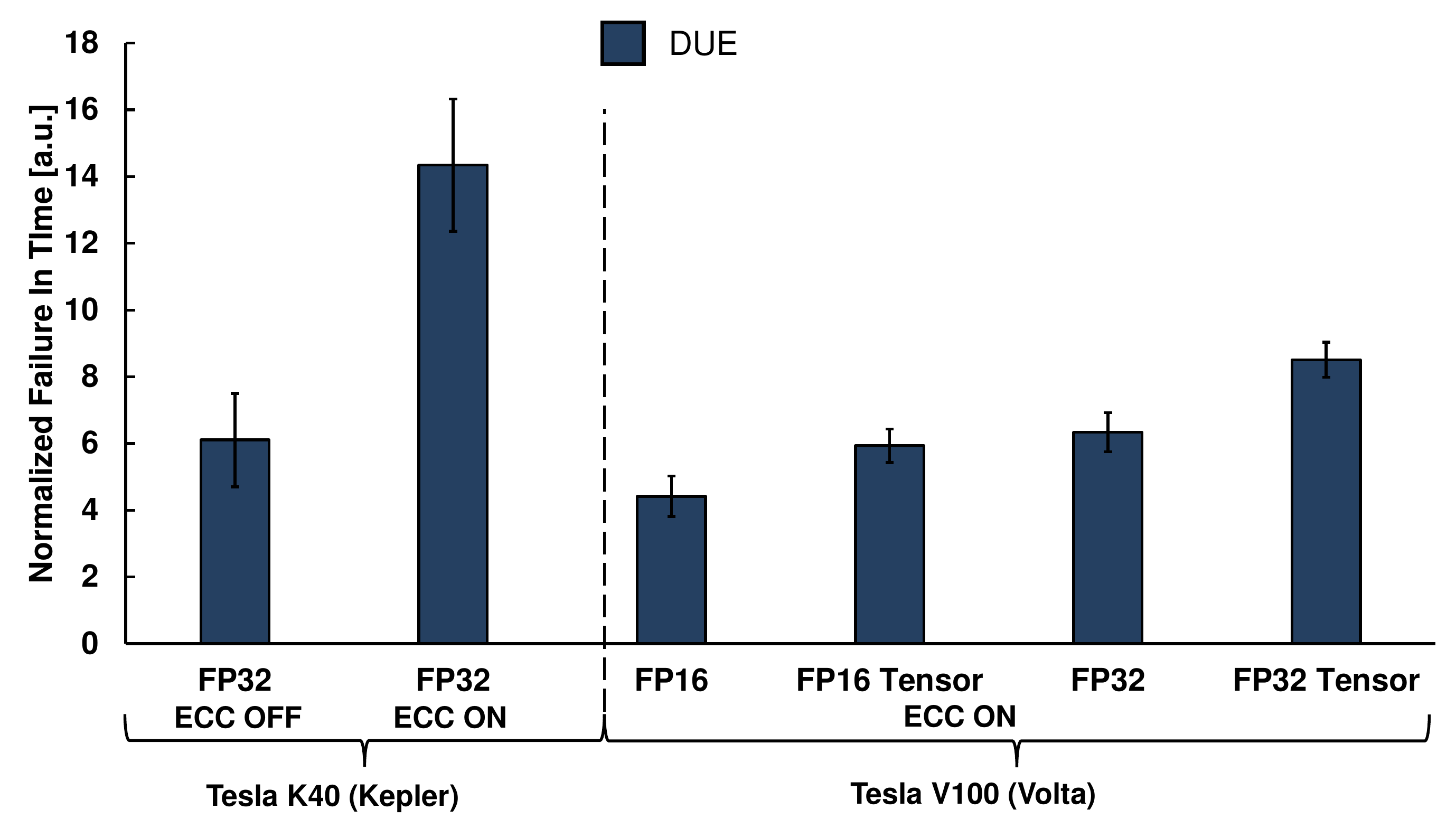}
        \label{fig:fit_gemm_due}
    }%
    \caption{{Figure~\ref{fig:fit_gemm_sdc} shows the normalized SDC Failure In Time (FiT) for the General Matrix Multiplication (GEMM) for Kepler and Volta. The SDC rate is separated by the error geometry observed on the output matrix. Figure~\ref{fig:fit_gemm_due} show the normalized DUE FiT for the GEMM on Kepler and Volta.}}%
    \label{fig:fit_gemm}
\end{figure*}

To completely characterize the critical errors on DNNs and how they can be generated at low level, we evaluated how the errors affect the General Matrix Multiplication algorithm (GEMM), used as the core of DNNs execution on modern GPUs. In fact, 70\% to 86\% of the operations for the evaluated DNNs are GEMM related operations.

{Figure~\ref{fig:fit_gemm} shows the SDC and DUE normalized FIT rates for GEMM}, respectively. Kepler values are relative to the SDC rate with ECC enabled, while Volta values are relative to the SDC rate of FP16 without Tensor cores. We selected matrix sizes that saturate the resources of each device ($2048\times{}2048$ for the Kepler and $16384\times{}16384$ for Tesla V100). As device resources are saturated, the data in Figure~\ref{fig:fit_gemm} can be used to compare the GEMM reliability across architectures. Smaller matrices will lower the FIT rate due to the unused resources. 

{
We also investigate the corrupted outputs of GEMM to classify the errors based on their coordinates (i.e., error geometry). Figure~\ref{fig:fit_gemm_sdc} shows the SDC rate separated into four types: (1) \textbf{Single} errors, only one corrupted element in the output matrix; (2) \textbf{Square}, four or more elements distributed in a rectangle; (3) \textbf{Line}, multiple corrupted elements on the same row/column of the output matrix; (4) \textbf{Random}, randomly distributed errors.
}
In most cases, GPU corruption affects more than a single output element. It is worth noting that multiple corrupted elements are not caused by multiple impinging particles. Instead, the impact of a single particle is \emph{spread} during computation across multiple output elements. We use this information to simulate multiple faults when performing fault injections at the SASS level (see Section~\ref{sec:fault_injection}).

{From Figure~\ref{fig:fit_gemm_due}, it is clear that GEMM reliability depends on both the device and the data/operation precision. The DUE rate when ECC OFF is lower than ECC ON on Kepler. When there is a double-bit error, ECC triggers an application DUE. When ECC is OFF, double-bit flips may lead to an SDC at the output of GEMM.}

We can notice that, independently of precision, the use of tensor cores increases DUE FIT. The GEMM with tensor cores is optimized to use the tensor cores to the maximum performance by performing Warp-level synchronizations. From the obtained results, the corruption of internal resources, necessary to perform specific Warp-level synchronizations of the tensor core, is particularly prone to generate a DUE. When the tensor cores are used to improve the performance of the DNNs, they may further increase the DNNs' DUE rate.

{Figure~\ref{fig:fit_gemm_sdc} shows that GEMM with tensor cores has a lower SDC FIT rate for all SDC types when executed in FP16 precision for Volta GPU}. This attests that the tensor core circuit is slightly more reliable than the combination of ADD, MUL, and the loop control variables needed to implement matrix multiplication in software. When GEMM is executed in FP32 precision, the tensor core SDC FIT rate increases significantly, being, {on average for the SDC types}, 20\% higher than the software GEMM FIT rate. 
In fact, as stated in Section~\ref{sec:methodology}, the tensor cores on Volta architecture execute physical operations only in FP16 precision. FP32 precision inputs require a hardware casting to FP16 precision, increasing the GPU's occupation and, thus, the SDCs (and DUEs) FIT rate. 
When executed on Volta without the tensor cores, the GEMM FIT rate increases as the precision increases. The increase from FP16 to FP32 is about $2\times{}$. As FP32 occupies $2\times{}$ more GPU resources, it is expected that the FIT rate will also have an increase with a factor close to $2\times{}$.

Square errors are potentially the most severe for DNNs. Square errors in GEMM are caused by faults that impact the scheduling or the execution of multiple threads in a Streaming Multiprocessor (SM). GEMM is a highly optimized version of matrix multiplication since it divides the input matrices into chunks (tiles) that fit in the cache of an SM, reducing memory latency. If a fault causes a thread to be incorrectly assigned or scheduled to an SM, or if some threads fail to synchronize, the whole SM output matrix part is likely to be corrupted, leading to a Square error. Errors in memory elements protected by ECC (e.g., registers and caches) manifest into either Single corrupted element or Line errors~\cite{Rech2013_Threads}. When ECC is turned ON, Single and Line errors (which are less critical for DNNs) are corrected, but the remaining errors (including rectangular errors) are not corrected. As a result, the percentage of Square errors increases when ECC is enabled.

{Figure~\ref{fig:fit_gemm_sdc} also shows that the geometry of the output errors slightly changes when comparing Kepler and Volta. For FP32 GEMM on both GPUs with ECC ON, the error distribution is mainly Square and Single errors}, showing that the criticality of the GEMM is directly related to how the error is propagated from the low-level fault to the algorithm output (details in Section~\ref{sec:fault_injection}).
It is worth noting that the criticality of the errors increases for the software GEMM compared to the GEMM optimized with the tensor core. The matrix multiplication performed in hardware generated more Single errors than the version only performed in software. This result is in line with past results~\cite{coralRubens2022}, where it has been shown that the reliability of DNNs hardware accelerators has a less critical error model than the operations performed only in software. 

In the next section, we present the fault injection results using different fault models on GEMM and YOLOv3. We demonstrate that using only the naive single-bit flip at the SASS level fault injection is not enough to mimic the errors observed in beam experiments.

%% file: src/faultinjection.tex
\section{Modeling complex faults with fault injection}
\label{sec:fault_injection}

\begin{figure*}[ht]%
	\centering
	\subfloat[]{
		\includegraphics[width=0.5\textwidth,keepaspectratio]{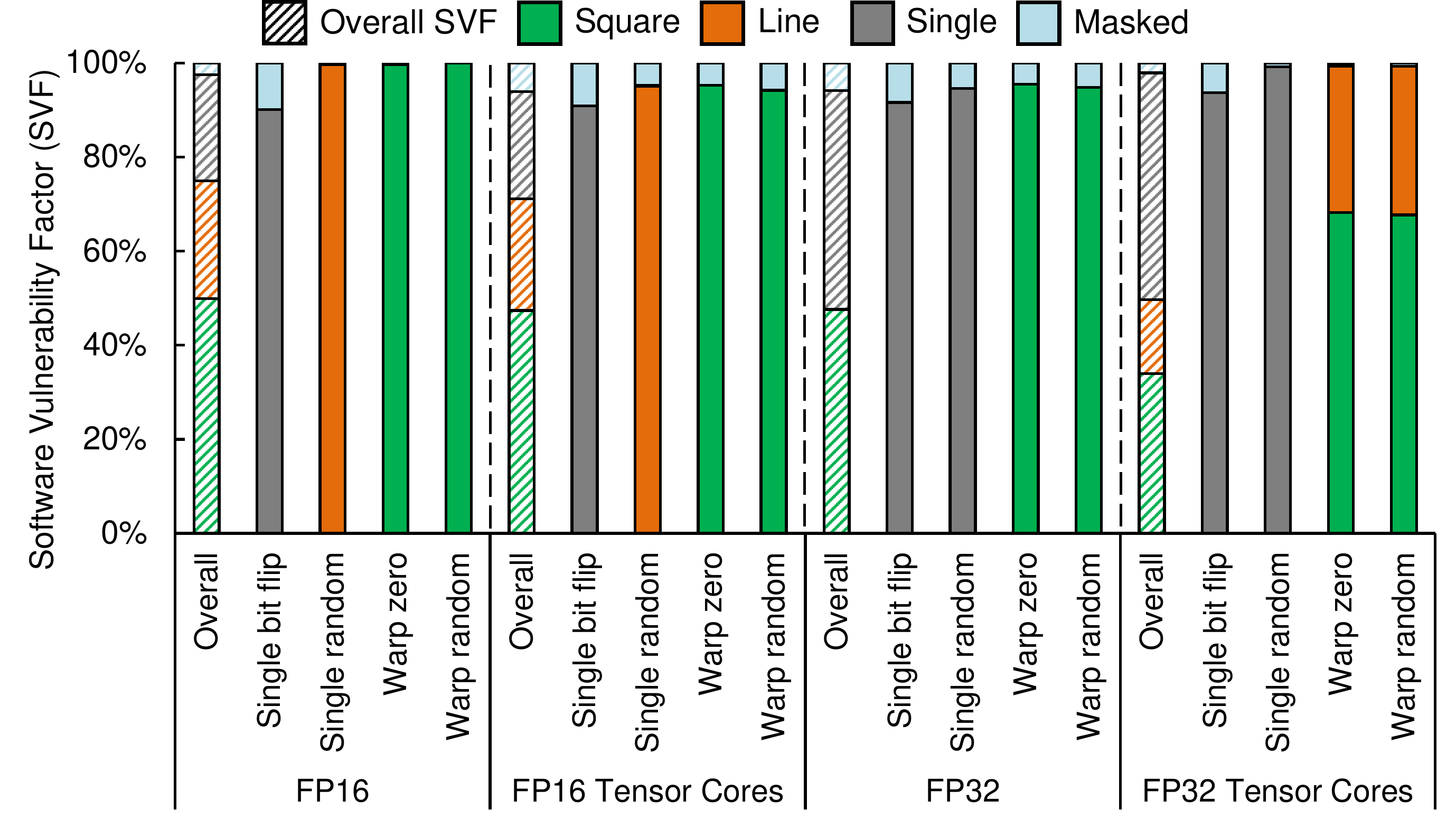}
		\label{fig:geometry_svf}
	}%
	\subfloat[]{
		\includegraphics[width=0.5\textwidth,keepaspectratio]{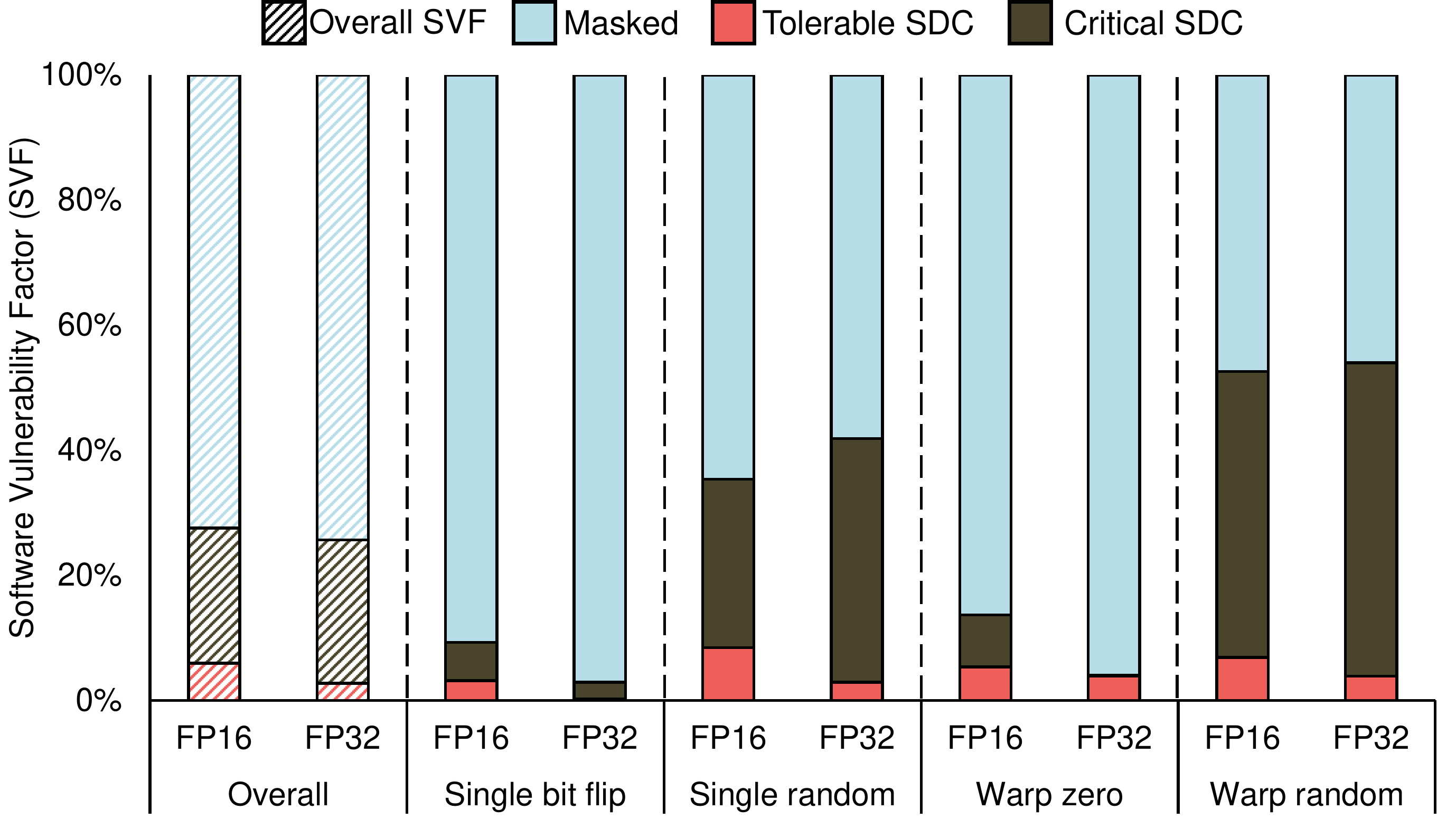}
		\label{fig:criticality_svf}
	}%
	\caption{Software Vulnerability Factor (SVF) for GEMM configurations on Figure~\ref{fig:geometry_svf} and YOLOv3 on Figure~\ref{fig:criticality_svf}. The outcome of each fault injection is grouped using the same methodology used for beam experiments results.}%
	\label{fig:svf}
\end{figure*}

This section presents the fault injection results with the multiple fault models proposed in Section~\ref{sec:background}. We show the Software Vulnerability Factor (SVF) for the injections performed on GEMM benchmarks for several configurations. To demonstrate the impact of the fault models on DNNs, we also show the YOLOv3 SVF for FP16 and FP32.

\subsection{Fault impact on GEMM}

Figure~\ref{fig:geometry_svf} shows the SVF for the fault injections performed on GEMM with Volta architecture, considering not only single-bit flips, but also more complex fault models. More precisely, we show results for four fault models, the standard Single Bit Flip, Single Random Value, Warp Zero Value, and Warp Random Value. The results are shown for two precisions, FP16 and FP32, with and without NVIDIA Volta tensor cores. The figure also presents the \textbf{Overall SVF}, obtained by combining all the fault models. Overall SVF serves as a reference to measure the impact of injecting a mix of faults instead of a single-bit fault model.

Results show that, despite the GEMM configuration, precision and algorithm, single-bit flip injections always produce single element corruption. However, as presented in Section~\ref{sec:fit_beam}, this model might not be realistic as an expected outcome from GEMM error criticality. 

It is worth noticing that when a single random value is injected, it produces different outcomes on FP16 and FP32 GEMM. In fact, this happens because of how FP16 instructions are computed on the Volta architecture. Each FP32 GPU core on Volta can execute two FP16 operations simultaneously, and, after the computation, the result is stored in a 32-bit register. Consequently, when an output register is replaced by a random or a zero value on FP16, it will always generate at least two incorrect elements. {
Injecting in the most/least significant bit of the 16-bit for FP16 would produce single-element injections. However, previous work with FP16 microbenchmarks with neutron beam experiments~\cite{ffsantos_hpca2019} has shown that the fault model of FP16 instructions when executing on FP32 cores leads to double element corruption at the 32 bits registers. Thus, our decision to modify the entire output of FP16 instructions aims to produce more realistic results.
}{
Additionally, as random and zero values always produce multiple errors on FP16 injections, the warp-wide injections on FP16 will, consequently, produce a predominance of Square errors.
}

Another exciting outcome of the complex fault model injections is the differences between the GEMM performed only on software (multiply and accumulate instructions) and using tensor core instructions.
For software-only GEMM, the outcomes of Warp-wide injections are 99\% and 95\% Square errors for FP16 and F32, respectively. However, when using tensor cores, 95\% and 69\% are Square errors for F16 and FP32, respectively. 
This is because when using tensor cores, the rounding mode is set to round towards zero~\cite{MarkidisTensorCoresRounding2018}, while the default operation is round to the nearest.
The rounding method can mask the fault before it is propagated to the output. Similar behavior is observed on beam experiments when comparing only software GEMM and GEMM with tensor cores.
{The fault propagation of the GEMM can also be impacted by the order of the instructions and cast operations on FP32 when using tensor cores. For each configuration on Figure~\ref{fig:geometry_svf}, CUBLAS calls a very optimized kernel, which may lead to a different fault masking. Additionally, when performing FP32 Tensor cores GEMM on Volta architectures, a cast from FP32/FP16 before and after the Tensor Cores instruction are necessary, which can lead to a difference in the fault propagation when comparing the outcomes of FP32 and FP16 with Tensor Cores.
}

When we compare the Overall SVF for the FP16 and FP32 configurations, the FP32 produces more single errors than line and square, while the contrary happens for FP16. Similarly, the same configurations on beam experiments that use FP32 produce more single errors than the FP16 ones. In fact, the errors with FP16 are more critical than the ones with FP32, as smaller errors can be more easily masked when using FP32.
It is worth noting that the high SVF produced on GEMM for complex fault models does not imply that they will generate only critical faults on a DNN. In the next section, we discuss the results of these fault models on YOLOv3. 

\subsection{Fault impact on YOLOv3}

Figure~\ref{fig:criticality_svf} shows the SVF for the injections on YOLOv3 object detection DNN for two floating-point precisions, FP16 and FP32, on Volta architecture. We use the same fault models as for the GEMM fault injections, i.e., Single Bit Flip, Single Random Value, Warp Zero Value, and Warp Random Value. The figure also shows the Overall SVF for both precisions.

The differences between FP16 and FP32 are lower for YOLOv3 than for GEMM (FP32 Overall SVF is 25.7\% while FP16 SVF is 27.6\%). As expected, the reduced precision can change the error rate of YOLOv3 but impacts less the fault propagation (SVF). 
Contrarily, the GPU fault model significantly impacts the criticality of YOLOv3 errors. The most critical errors on YOLOv3 come from Single Random Value and Warp Random Value, 26.9\% and 45.7\% of critical SDCs with FP16, respectively, and 39.0\% and 50.1\% of critical SDCs with FP32, respectively. Similar behavior has been demonstrated with security attacks on DNNs, where the corruption of a single float exponent is enough to degrade the accuracy of some DNNs by 69\%~\cite{RakinBitFlipAttack2019, LiDefendingAttack2020}. Consequently, it is expected that the \textit{Random Value} based fault models will produce more critical errors.

An exciting result from Figure~\ref{fig:criticality_svf} is the impact of the single-bit flip and Warp zero value fault models on YOLOv3. It is known from the GEMM fault injections that single-bit flip injections will generate single errors. Moreover, the single-bit flip on the lower places of the floating-point representation will generate small perturbations on YOLOv3 inference, leading to a low SVF for FP16 and FP32 precisions. The Single bit flip fault model produces an SVF of 9.3\% for FP16 and 2.9\% for FP32, and the Warp zero value produces an SVF of 13.7\% for FP16 and 4.0\% for FPF32.
Correspondingly, zeroing DNN parameters can act more like a regularizer instead of reducing the accuracy of the network~\cite{Hinton2012Dropout, ShangRelu2016}. 
The SVF for YOLOv3, when Warp Zero Value is injected, is 2.0$\times$ lower than the Overall SVF for FP16 and 6.4$\times$  lower than the Overall SVF for FP32.

\begin{figure}[t]
	\centering{
		\includegraphics[width=\linewidth]{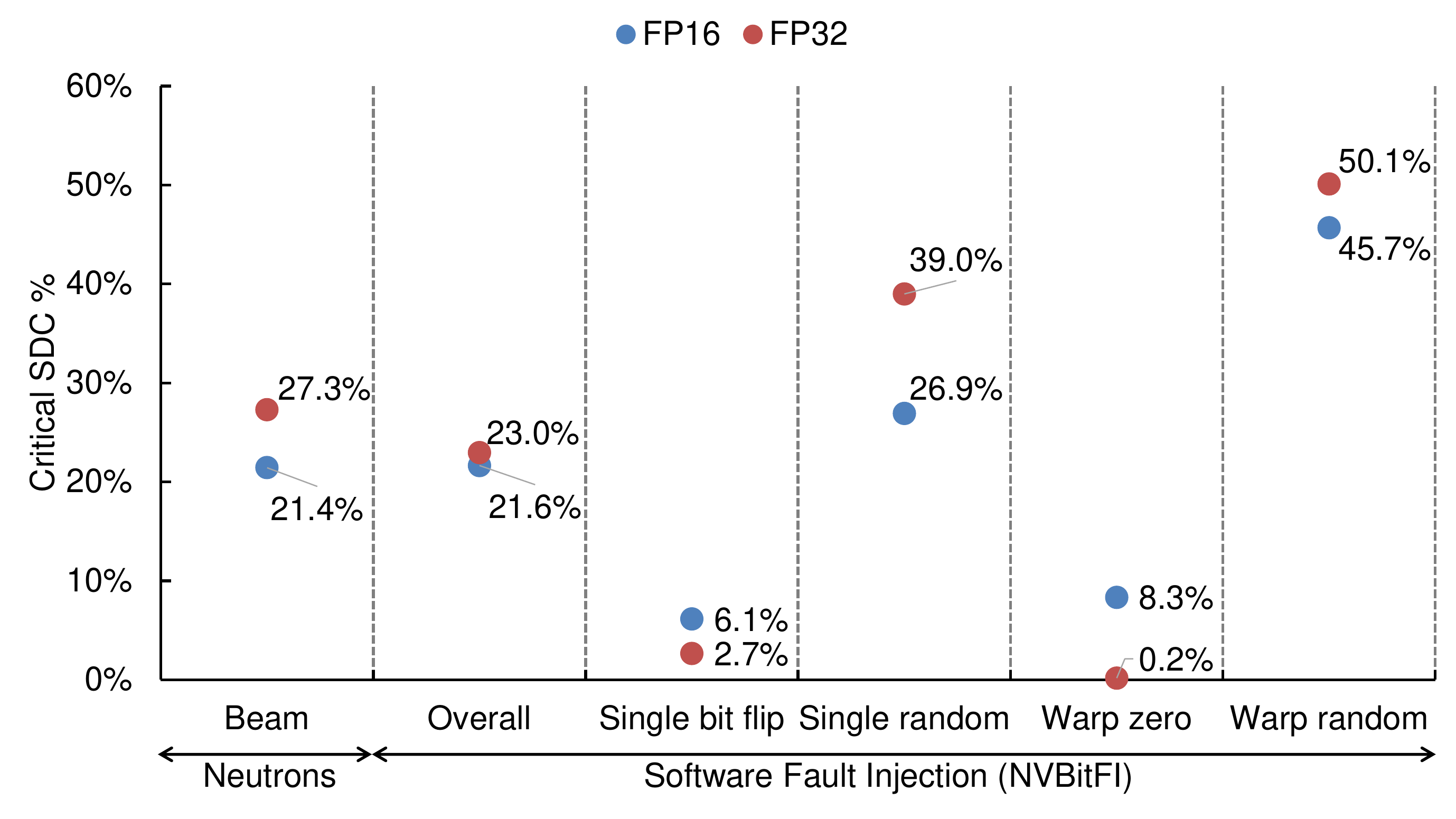}
		\caption[]{Comparison between the YOLOv3 SDC criticality on neutron beam experiments vs. NVBitFI.}
		\label{fig:beam_vs_fi}
	}
\end{figure}

Assuming that only one fault model is sufficient to produce a realistic DNN reliability evaluation is an oversimplification, since various types of faults are generated from the neutron iterations with the hardware in the beam experiments. 
To validate our findings, Figure~\ref{fig:beam_vs_fi} compares the critical SDCs rate measured with neutron beam experiments and with fault injection.
As shown, when a cocktail of different fault models is injected, we observe a percentage of critical SDCs which is very similar to the beam experiments.
The Critical SDCs of Overall SVF are 47\% and 46\% lower than the most critical fault model (Warp Random Value) for FP16 and FP32, respectively. The Critical SDCs of Overall SVF are 3.53$\times$ and 8.66$\times$ higher than Single bit flip for FP16 and FP32, respectively.
These results confirm that injecting single-bit flip faults at the software level is unrealistic for the reliability evaluation of DNNs on GPUs, as it can underestimate the real SDC criticality.

Finally, we recall that our presented fault models are based on syndromes observed on RTL fault injections which we propagate to the SASS-level to explain the error criticality observed in neutron beam experiments. We are able to quickly simulate the sources of critical errors in software without having to employ fault injections using RTL or microarchitectural simulations. Lower level fault injections are in fact more accurate, but very slow. A simple GEMM evaluation for one configuration as presented in Figure~\ref{fig:geometry_svf} would take $4.8\times10^{5}$ hours to simulate according to recent works~\cite{condia2020}. This would make the reliability evaluation of DNNs unfeasible. 
%
We acknowledge that our approach presents some limitations as the fault injection is based on SASS-level instrumentation, that is, software level. Consequently, our fault injection is limited to the user-accessible resources, making it impossible to simulate faults that affect resources like memory buffers.
However, despite the proposed fault models' limitations, we can still simulate the syndrome of a fault in a shared resource of GPU faults at the SASS level.


%% file: src/conclusion.tex
\section{Conclusions}
\label{sec:conclusions}

We have discussed the reliability of DNNs on beam experiments and evaluated the criticality of the errors in the output. 
Although the ECC is a powerful technique to reduce the SDC rate, it cannot reduce the Critical errors proportionally when DNNs are considered as an application. 
We then characterized the GEMM kernels as the core of DNNs, to investigate the error models that can impact the DNNs reliability. 
As radiation experiments demonstrated, the single bit flip in the memories is not the leading cause of critical errors. Additionally, the criticality of the output of the GEMM indicates that a single error in a single element of a GEMM kernel is not realistic as an error model to simulate critical errors on DNNs. 
We then have performed a SASS-level fault injection on YOLOv3 object detection with different fault models to investigate the leading cause of the critical errors. We show that it is possible to simulate complex fault models at a higher level of fault injection, and that injecting single-bit flips at instruction output may not be the most realistic fault model for complex applications like DNNs.